%% file: ver15.tex
\documentclass[psfig]{mn2e}
\usepackage{graphicx}

\footnotesize
\newdimen\digitwidth    
\setbox0=\hbox{\rm0} \digitwidth=\wd0 \catcode`!=\active
\def!{\kern\digitwidth}
\normalsize

\newcommand\nuddd{\ifmmode\stackrel{\bf \,...}{\textstyle \nu}\else$\stackrel{\,...}{\textstyle \nu}$\fi}
\def\lsim{~\rlap{$<$}{\lower 1.0ex\hbox{$\sim$}}}

\title{Timing measurements and proper motions of 74 pulsars using the Nanshan radio telescope}

\author[Zou, Hobbs et al.]{W. Z. Zou,$^{1,2}$\thanks{Email: zouwz@ms.xjb.ac.cn}
G. Hobbs,$^{3}$ ~~ N. Wang,$^{1}$ ~~R. N. Manchester$^{3}$
\newauthor X. J. Wu,$^{4}$~~H. X. Wang,$^{1}$\\
$^{1}$ National Astronomical Observatories, CAS, 40 South
Beijing Road, Urumqi, 830011, China\\
$^{2}$ Graduate School of CAS, 19 Yuquan Road, Beijing, 100039, China\\
$^{3}$ Australia Telescope National Facility, CSIRO, PO Box 76,
Epping, NSW 1710, Australia\\
$^{4}$ Astronomy Department, School of Physics, Peking University,
Beijing, 100871, China\\}

\begin{document}
\maketitle
\pagestyle{plain}

\begin{abstract}

We have measured the positions of 74 pulsars from regular timing
observations using the Nanshan radio telescope at Urumqi
Observatory between 2000 January and 2004 August (MJD 51500 --
53240). Proper motions were determined for these pulsars by
comparing their current positions with positions given in pulsar
catalogues. We compare our results to earlier measurements in the
literature and show that, in general, the values agree. New or
improved proper motions are obtained for 16 pulsars. The effect of
period fluctuations and other timing noise on the determination of
pulsar positions is investigated. For our sample, the mean and rms
transverse velocities are 443 and 224 km~s$^{-1}$ respectively,
agreeing with previous work even though we determine distances
using the new NE2001 electron density model.
\end{abstract}


\section{Introduction}

Most pulsar proper motions have been obtained using interferometric
observations \cite{las82,hla93,fgml97,bfg+03}. However, pulsar timing
also allows measurement of very accurate pulsar positions and hence
proper motions \cite{mtv74,tsb+99}. This method was considered to
produce inaccurate results for young pulsars whose timing residuals
may be dominated by glitches and timing noise.  Recently, Hobbs et
al. (2004)\nocite{hlk+04} showed that with data spanning more than six
years it is possible to model the timing noise by fitting harmonically
related sinusoids. The spectrum of this timing noise has little power
at periods corresponding to one year and so any timing noise can be
removed from the timing residuals whilst leaving the signatures of an
incorrect position determination or unmodelled proper motion.  In this
paper, we use the simple method of determining pulsar proper motions
by comparing current positions obtained with nearly five years of
timing data from the Nanshan radio telescope at Urumqi Observatory in
China to earlier catalogued values.  Having only five years of
observations means that it is possible to fit just a few harmonically
related sinusoids to the data, which in many cases leaves significant
structures in the timing residuals after the whitening procedure. We
therefore choose to whiten the residuals using polynomials instead of
sinusoids. We compare the results from this analysis with published
measurements of the proper motions and consider the effects of timing
noise in these relatively short data spans.

Proper motion measurements have shown that pulsars are
high-velocity celestial objects with typical velocities of several
hundred km s$^{-1}$ \cite{ll94}, but it is not clear how the
pulsars achieve such large velocities.  The exact form of this
velocity distribution is also not well understood with
Arzoumanian, Chernoff \& Cordes (2002)\nocite{acc02} finding
evidence for a bimodal distribution, but Hobbs et al. (2005)
finding no such evidence. An accurate determination of the proper
motions is therefore essential in understanding the velocity
distribution.  However, the literature contains many examples of
inconsistent proper motion and position measurements. For instance
Lyne, Anderson \& Salter (1982)\nocite{las82} obtained $\mu_\alpha
= -102(5)$\,mas\,yr$^{-1}$ for the proper motion in right
ascension\footnote{Uncertainties given in parentheses after a
quantity refer to the last quoted digit and are $1\sigma$
values.} for PSR~B1133$+$16 whereas Brisken (2001)\nocite{bri01}
published that $\mu_\alpha = -74.0(4)$\,mas\,yr$^{-1}$. Similarly
Hobbs et al. (2004) determined a proper motion of $\mu_\alpha =
-10(12)$\,mas\,yr$^{-1}$ for B1552$-$31 that is inconsistent with
the value of 61(19)\,mas\,yr$^{-1}$ given by Brisken et al.
(2002)\nocite{bbgt02}. It is therefore essential to provide many
checks on the correctness of existing proper motion values as well
as providing new or improved values.

In \S 2, we briefly describe the observing system at the Nanshan radio
telescope.  In \S 3 we present our analysis method and the derived
positions and proper motion values.  The discussion in \S 4 is divided
into i) an analysis of the effect of timing noise on position
determinations, ii) a comparison of our proper motion results with
earlier published values, iii) the velocities obtained using our
complete sample of proper motions and iv) new and improved proper motions
from our work.

\section{Observations and Analysis}

Timing observations of 74 pulsars using the Nanshan 25-m radio
telescope operated by the Urumqi Observatory, National
Astronomical Observatories of China, have been regularly carried
out from 2000 January to 2002 June using a room temperature
receiver at 1540\,MHz. Since 2002 July, 284 pulsars have been
monitored approximately once every nine days using a dual-channel
cryogenic system that receives orthogonal linear polarisations at
the central observing frequency of 1540\,MHz. The pulsars were
chosen to be observable from the Observatory (which has a lower
declination limit of $-43\degr$) and to have mean flux densities
greater than 0.5~mJy. In this paper, we consider results for the
74 pulsars with mean flux densities larger than 2~mJy and
distances within  8.5~kpc over the data span 2000 January to 2004
August.

After down-conversion, the two polarisations are each fed to a
filterbank consisting of 128 contiguous channels each of width
2.5\,MHz.  The signals from each channel are square-law detected,
filtered and one-bit sampled at 1~ms intervals using a data
acquisition system based on a PC operating under Windows NT (full
details of the observing system are provided in Wang et
al. 2001).\nocite{wmz+01} The data are folded synchronously at the
pulsar topocentric period for sub-integration times of between one and
four minutes and written to disk. Time-tagging of the pulse
integrations is provided by a hydrogen maser clock calibrated using
the Global Positioning System (GPS) signals and a latched microsecond
counter. Off-line programs sum the orthogonal polarisations and
dedisperse the data into eight sub-bands. Observations of a given
pulsar typically consist of four sub-integrations with observation
lengths of between 4 and 16 minutes depending on the flux density of
the pulsar.  The folded profiles are stored on disk for subsequent
processing and analysis.

We use the {\sc psrchive} software package \cite{hvm04} to obtain
the pulse times of arrival (TOAs) from each observation. To do
this, the pulse profiles from each integration are
cross-correlated with high signal-to-noise templates to produce
topocentric TOAs.  These are subsequently processed using the
standard timing program {\sc tempo}\footnote{See
http://www.atnf.csiro.au/research/pulsar/tempo} with the improved
Jet Propulsion Laboratory ephemeris DE405 \footnote{See
http://ssd.jpl.nasa.gov/iau-comm4/relateds.html}. TOAs are
weighted by the inverse square of their uncertainty, with these
being scaled (if necessary) so that the fitting routine gives
reduced $\chi^2$ values close to 1.0.  Uncertainties in the fitted
parameters for most pulsars are taken to be twice the standard
errors obtained from {\sc tempo}.  However, for pulsars whose
timing residuals are dominated by timing noise the uncertainties
in the fitted positions are taken to be three times the formal
\textsc{tempo} errors. These error estimates are discussed in more
detail in \S 4.

\section{Results}

\begin{figure*}
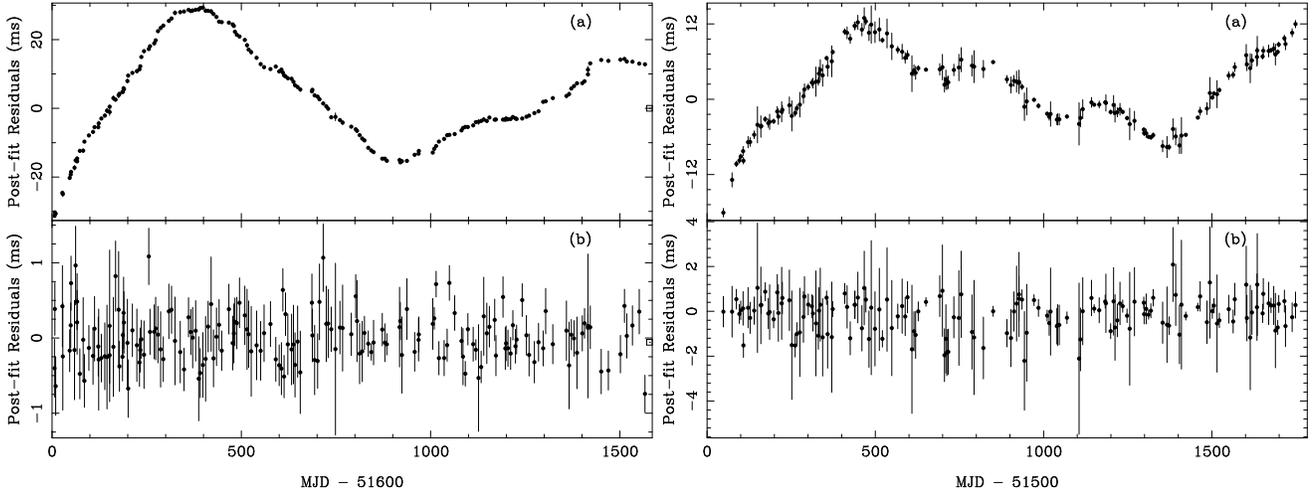

\centering
\includegraphics[width=7.0cm,angle=-90]{J0139.ps}
\includegraphics[width=7.0cm,angle=-90]{J1709.ps}
\caption{The post-fit timing residuals for PSRs J0139+5814 and J1709$-$1640
(a) after fitting for the rotational frequency and its first derivative and
(b) after fitting multiple frequency derivatives.}
\label{fg:timing-noise}
\end{figure*}

In this paper we present the results from an analysis of TOA data
for 74 pulsars. These results include new positions from which we
determine proper motions and velocities. For each of the 74
pulsars, the position was measured by setting the reference epoch
of the timing model to an integral MJD near the centre of the data
span. We fit the pulsar position, rotational frequency and its
first derivative. For pulsars whose timing residuals are dominated
by timing noise we fit multiple frequency derivatives to model and
hence effectively remove the timing noise.
For instance, Figure~\ref{fg:timing-noise} plots the post-fit timing residuals
for PSRs~J0139+5814 and J1709$-$1640 after fitting for the rotational
frequency and its first derivative and after fitting multiple frequency
derivatives. Among the 74 pulsars,
the residuals of 26 are dominated by timing noise and ten glitches
in total were found in five pulsars. Some timing results were
described by Zou et al. (2004)\nocite{zww+04} and others will be
reported in a subsequent paper. The typical post-fit residuals of
the 74 pulsars are several hundred $\mu$s. All of the pulsars in
our sample are solitary and none are millisecond pulsars. Most
have ages in the range $10^4 - 10^7$~yr and all are within 9~kpc
of the Sun.

In order to obtain proper motions we compared our results with early
previously published position estimates.  However, the very earliest
position estimates often used relatively poor instrumentation and so
can have large uncertainties. For each pulsar, we therefore select the
earliest available position estimate that has an uncertainty less than
3\arcsec.  For the majority of our pulsars, we obtained these
positions from the Taylor, Manchester \& Lyne (1993)\nocite{tml93}
catalogue.  For PSRs~B0525$+$21, B1749$-$28, B1754$-$24 and B1821$-$19
we use positions given by Hobbs et al. (2004). For most pulsars, the
epoch of the positions given by the catalogue is more than ten years
earlier than the epoch for our data.  The maximum time span is 33~yr.

\input{table1}

\input{table2}

Table 1 gives the pulsar names in J2000 coordinates and B1950
coordinates in the first two columns. The following five columns give,
respectively, the epoch of the position measurement, the right
ascension and declination from the timing analysis, the number of TOAs
in the nearly five-year data set and the time span between the two
position determinations.  The final four columns refer to the
previously published position and give the epoch, right ascension,
declination and a reference for the measurement.

Proper motions derived by comparison of the Nanshan timing results
with previously published positions for the 74 pulsars as well as the
most reliable previously published timing and interferometric proper
motions for these pulsars are given in Table 2. For most of the
pulsars in our sample we determine the pulsar distance $D$, given in
the third column of the table, using the dispersion measure and the
Cordes \& Lazio (2002; hereafter NE2001)\nocite{cl02} electron density
model. The exceptions are 13 pulsars for which interferometric annual
parallax measurements are available\footnote{ATNF Pulsar Catalogue\\
http://www.atnf.csiro.au/research/pulsar/psrcat}. The next three
columns give the proper motion in the right-ascension direction
($\mu_\alpha^c$) the declination direction ($\mu_\delta^c$) and
the total proper motion ($\mu_{tot}^c$) derived from the data
listed in Table 1. If proper motions already exist in the
literature, we also provide the most reliable proper motions
available from timing (superscript T) and from interferometry
(superscript I).  All of the timing measurements are from Hobbs et
al. (2004) except that for PSR~J0738$-$4042 which is from Siegman
et al. (1993)\nocite{smd93}. References for the interferometric
proper motions are given in the 11th column. Offsets of
$\mu_\alpha^c$ and $\mu_\delta^c$ from the most accurate of the
previously published proper motions normalised by the uncertainty
in $\mu_\alpha^c$ and $\mu_\delta^c$, respectively, are given in
the next two columns ($\epsilon_\alpha$ and $\epsilon_\delta$). In
the final column, values of the pulsar transverse velocity derived
from $\mu_{tot}^c$ and the distance.

\begin{figure*}
\includegraphics[width=6cm,angle=-90]{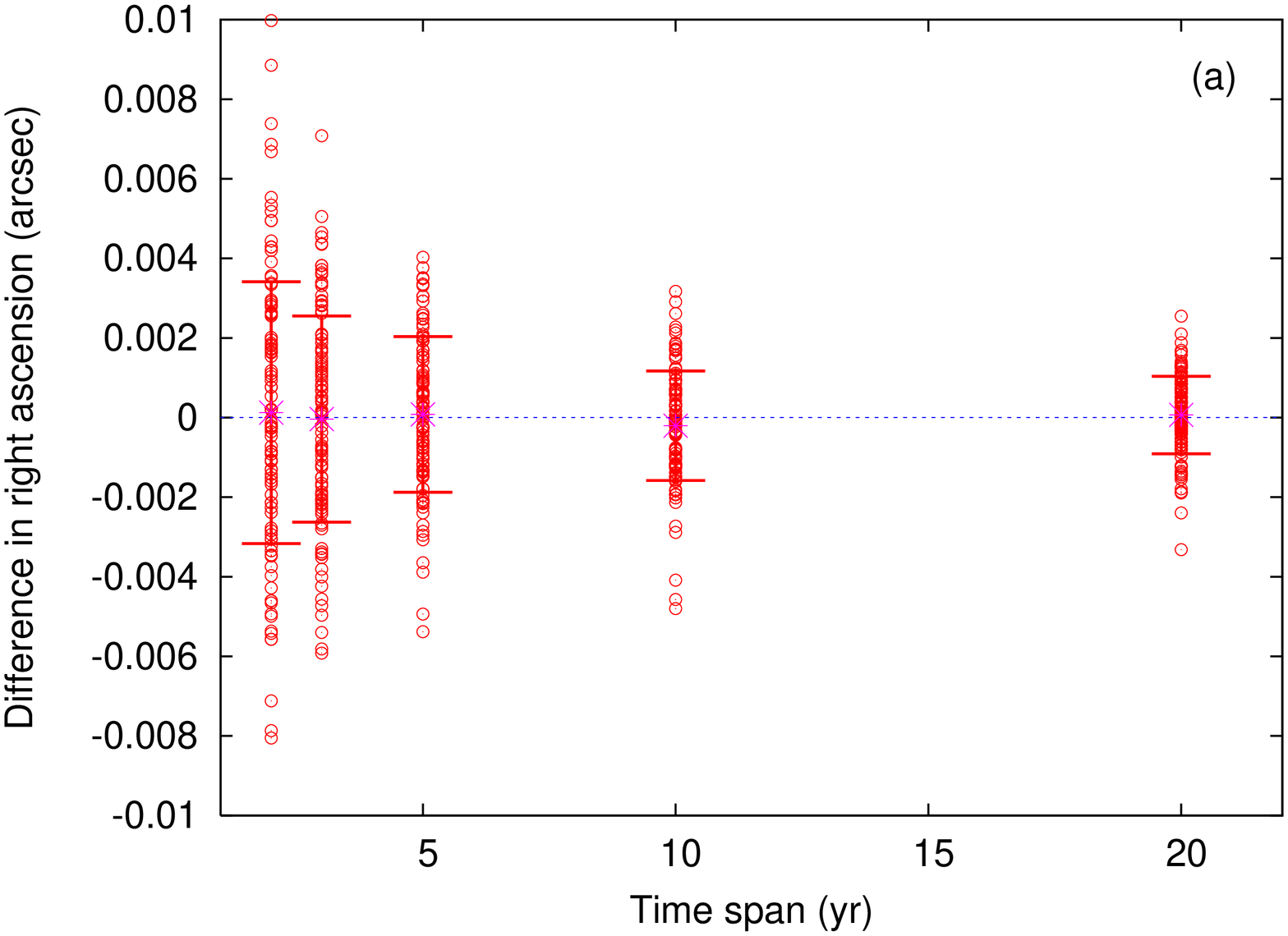}
\includegraphics[width=6cm,angle=-90]{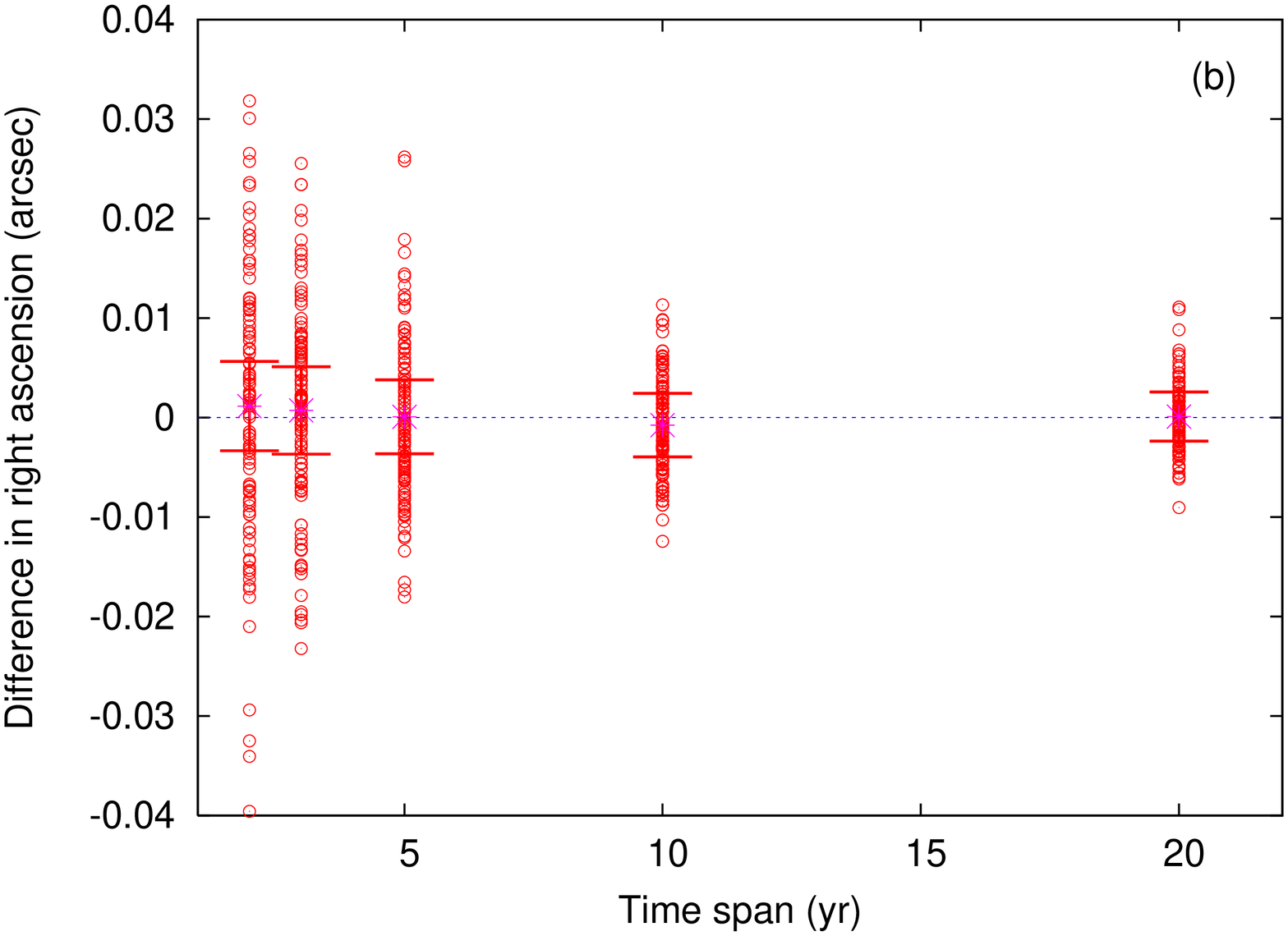}
\caption{Distribution of offsets in right ascension obtained from
   a timing fit to 100 simulated data sets
   dominated by (a) white noise and (b) red noise for time spans of 2,
   3, 5, 10 and 20 years.  The error bars are the mean formal error
   given by {\sc tempo} for each of the data spans; see text.}\label{fg:simulate}
\end{figure*}

\section{Discussion}

\subsection{Errors in position measurements based on pulsar timing}

The precision and reliability of proper motions determined by
comparison of timing positions with other positions are clearly
affected by the presence of timing noise. In this section we
attempt to quantify the uncertainty in positions derived from
pulsar timing as a function of the amplitude and spectrum of the
timing noise and the length of the data span. Timing noise
consists of two components: a flat-spectrum or `white' component,
usually dominated by receiver noise, and a steep-spectrum or `red'
component normally dominated by fluctuations in the pulsar spin
rate. The red or intrinsic component is generally larger in young
pulsars and is often quantified by the stability parameter as
\begin{equation}
\Delta_8\equiv \log\left(\frac{1}{6\nu}|\ddot\nu|t^3\right),
 \label{eq:stability}
\end{equation}
where $\nu = 1/P$ and $\ddot\nu$ is its second time derivative and
$t=10^8$~s~\cite{antt94}. These authors show that $\Delta_8$ is
approximately proportional to $\dot P^{1/2}$, where $\dot P$ is the
first time-derivative of the pulsar period. Timing noise can also be
characterised by the power spectrum of the timing residuals $S(\nu) \sim
\nu^\alpha$, where $\alpha = -2$ if the noise results from a random
walk in pulse phase, $\alpha = -4$ for frequency noise and $\alpha =
-6$ for slowing down noise (Boynton et al. 1972; Groth
1975)\nocite{bgh+72,gro75b}. However, more recent work \cite{ddm97}
has shown that the power-law spectra are often complex with several
power-law segments.

We investigate the effect of timing noise on position determinations
by simulating pulse TOAs containing a mixture of white and red noise
using the {\sc tempo2} software package \cite{hem05}. The red noise
has a spectral index $\alpha = -3$ chosen to give an approximate
representation of the timing noise seen in our residuals ($\alpha = 0$ for white noise).
TOAs were computed for data spans ($T$) of 2, 3, 5, 10 and 20\,years at 10-day
intervals and adjusted so that the residuals matched the desired noise
model. This process was repeated to give 100 sets of TOAs for each of
the five data spans.  Figure~\ref{fg:simulate} shows the distribution
of offsets of the right ascension obtained from the timing fit
compared to the nominal value for data sets dominated by white noise
(a) and red noise (b). Figure~\ref{fg:simulate}a shows that for white
noise the error in the timing position decreases with increasing data
span, approximately proportional to $T^{-1/2}$ as
expected. Furthermore, the standard deviation given by {\sc tempo}
accurately represents the uncertainty, with 68\% of offsets within
$\pm 1\sigma$. Figure~\ref{fg:simulate}b shows that, for red noise,
the {\sc tempo} standard deviation greatly under-estimates the actual
scatter, especially for short data spans, and that it doesn't decrease
as expected with increasing data span.  For a timespan of 5\,years,
only 27\% of the offsets are within $\pm 1\sigma$ of the {\sc tempo}
solution.

\begin{figure}
 \includegraphics[width=6cm,angle=270]{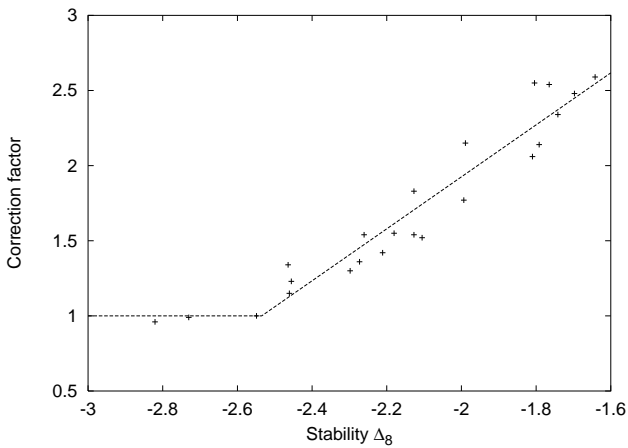}
 \caption{The multiplicative factor required for the \textsc{tempo}
 uncertainties to equal the true uncertainties for different amounts
 of timing noise.  Each point is obtained from 100 simulations of 5\,year
 data sets with TOA uncertainties of 0.5\,ms.}\label{fg:errFactor}
\end{figure}

For the five-year data span relevant to the present observations, we
plot in Figure~\ref{fg:errFactor} an error factor, defined as the rms
scatter of the points in Figure~\ref{fg:simulate}b divided by the {\sc
tempo} standard deviation, versus the $\Delta_8$ parameter. Typically,
the range of $\Delta_8$ represented by the 74 pulsars in the sample is
between $-3.8$ and 0, and the mean value of $\Delta_8$ is $-2.2$.  The
precise shape of the error factor curve depends on the whitening
procedure, the power-law slope, the typical TOA uncertainties and
systematic effects that may exist in the real data.

\begin{figure*}
\includegraphics[width=6cm,angle=270]{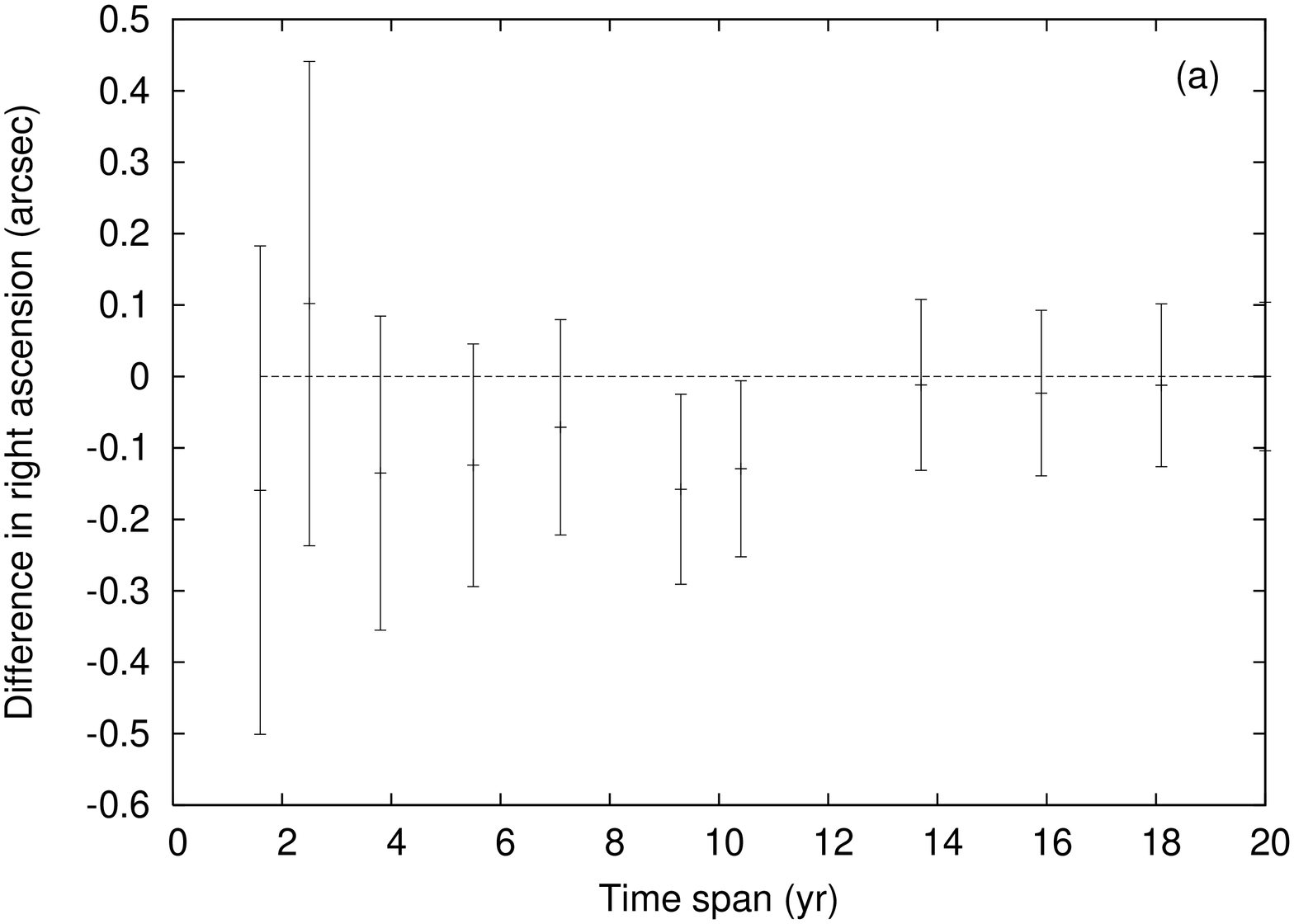}
\includegraphics[width=6cm,angle=270]{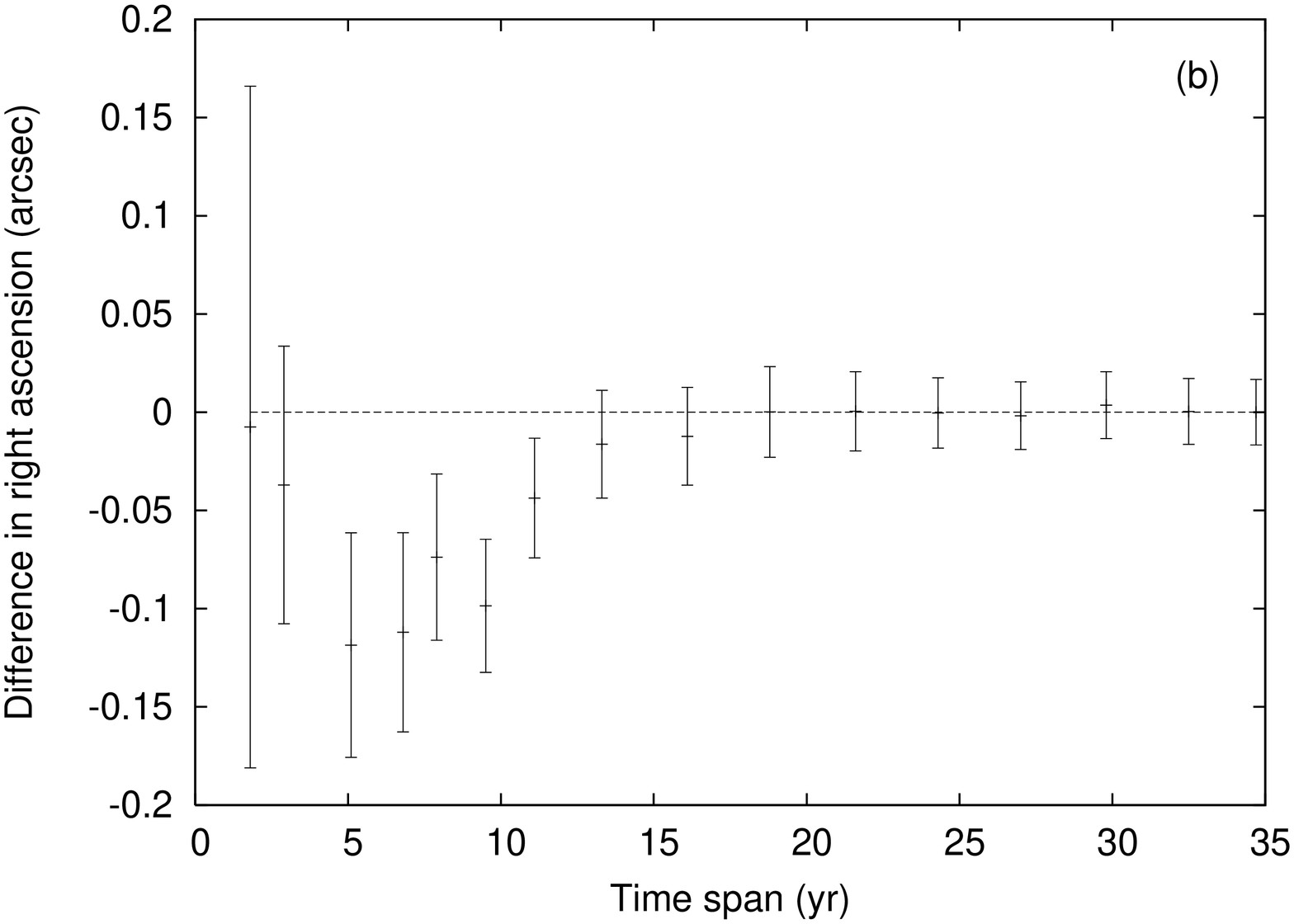}
\caption{Right ascension offsets and their errors estimated using
{\sc tempo} for data sets of different lengths for (a) PSR~B0320+39 and (b)
PSR~B1133+16. Data are from Jodrell Bank Observatory. }\label{fg:real}
\end{figure*}

To confirm the results of these simulations, we analysed data for two
pulsars with long data spans supplied to us from Jodrell Bank
Observatory. Details of the data acquisition and analysis systems are
given by Hobbs et al. (2004). We first analyse a 20-year data set for
PSR~B0320$+$39 which contains little or no red timing noise.  We
select different data spans that are centred on MJD~49290 and fit for
the pulsar position, setting the proper motion to the interferometric
value. Right ascension offsets relative to the value from the full
data span and their uncertainties as estimated by {\sc tempo} are
shown in Figure~\ref{fg:real}a. Results of a similar analysis using a
35-year data set for PSR~B1133$+$16, which shows a moderate amount of
timing noise, are shown in Figure~\ref{fg:real}b. In line with common
practice, higher-order frequency-derivative terms were used to whiten
(i.e., absorb the red noise) from the final residuals when fitting for
the position. Figure~\ref{fg:real}a confirms that {\sc tempo} standard
deviations are realistic error estimates for a pulsar with little or
no red timing noise for either short or long data
spans. Figure~\ref{fg:real}b further illustrates that for noisy
pulsars, data spans in excess of about 20 years are necessary for
reliable position estimates and that {\sc tempo} errors underestimate
the true uncertainty by about a factor of two for data spans of the
order of five years.

Consistent with these results, for the Nanshan timing positions in
Table 1 we have adopted an uncertainty of twice the formal {\sc tempo}
standard deviation for pulsars that show little timing noise
($\Delta_8 < -1.8$), whereas for pulsars with $\Delta_8 > -1.8$ we
multiply the formal errors by three.

\subsection{Comparison of derived proper motions}

\begin{figure*}
\includegraphics[width=8.5cm,angle=0]{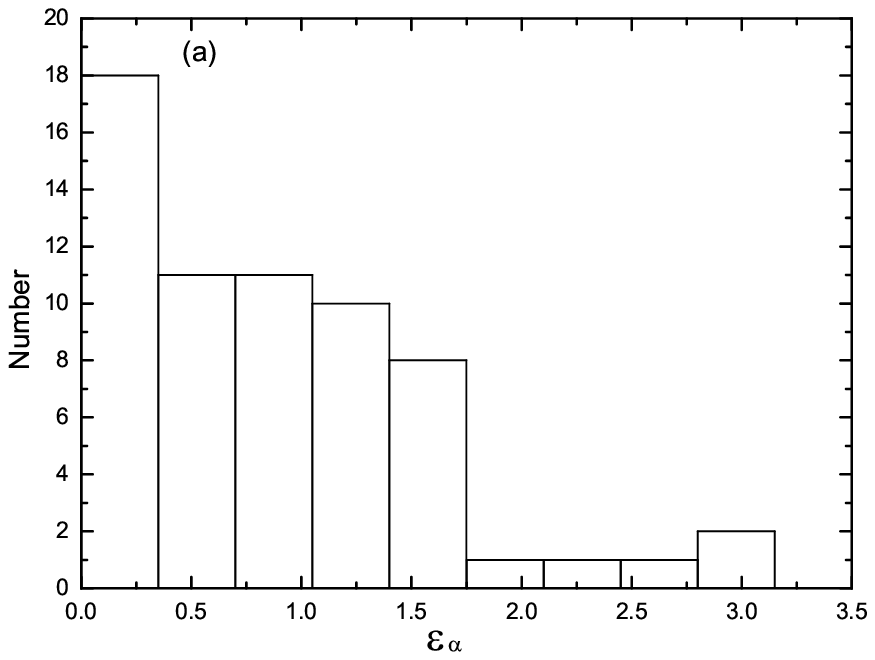}
\includegraphics[width=8.5cm,angle=0]{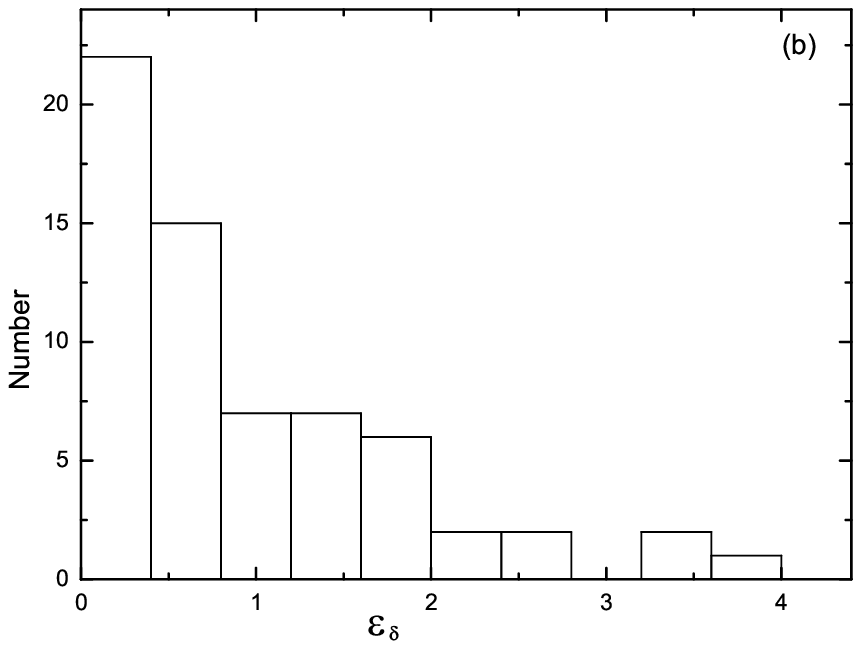}
\caption{Distributions of offsets of derived proper motions from the
   best independently measured values relative to the estimated
   uncertainties (a) in right ascension and (b) in declination. PSR~B0736$-$40 has anomalously large offsets and is
   excluded from the plots.}\label{fg:epsHist}
\end{figure*}

Values of $\epsilon_\alpha$ and $\epsilon_\delta$ given in Table 2
show that, with a few exceptions, the proper motions derived by
comparison of Nanshan timing positions with previously catalogued
positions agree well with the best independently derived proper
motions. This is further illustrated in Figure~\ref{fg:epsHist} which
shows the distribution of (a) $\epsilon_\alpha$ and (b) $\epsilon_\delta$
values. These plots exclude PSR~B0736$-$40; see below for a
discussion of this pulsar. These distributions are close to Gaussian
-- 63\% of the derived proper motions in right ascension lie within
1$\sigma$, 92\% within 2$\sigma$ and 99\% within 3$\sigma$ of the
independent results -- showing that error estimates are realistic.
Similar results are obtained for declination where 63\% are within
1$\sigma$, 92\% within 2$\sigma$ and 97\% within 3$\sigma$.

However, there are a few conspicuous outlying points. For
PSR~B0736$-$40, the scaled offsets $\epsilon_\alpha$ and
$\epsilon_\delta$ are very large, 11.9 and 5.5 respectively (Table
2). To derive the proper motion, the Nanshan timing position was
compared with the timing position given by Downs \& Reichley
(1983)\nocite{dr83} to give $\mu_\alpha = -48(2)$ and $\mu_\delta =
35(2)$~mas~yr$^{-1}$. The $\epsilon$ values are based on comparison of
these proper motions with the Brisken et al. (2003) interferometric
proper motions given in Table 2: $\mu_\alpha = -14.0(12)$ and
$\mu_\delta = 13(2)$~mas~yr$^{-1}$. Downs \& Reichley (1983)
independently determined a proper motion from their timing data:
$\mu_\alpha = -56(9)$ and $\mu_\delta = 46(8)$~mas~yr$^{-1}$, and
Siegman, Manchester \& Durdin (1993)\nocite{smd93} compared their
Molonglo timing position with the Downs \& Reichley (1983) position to
obtain $\mu_\alpha = -60(10)$ and $\mu_\delta = 40(10)$~mas~yr$^{-1}$
(Table 2). For this pulsar it is possible to solve for a proper motion
using the Nanshan timing data alone, giving $\mu_\alpha = -51(22)$ and
$\mu_\delta = 6(24)$~mas~yr$^{-1}$. Although not fully independent,
all of these timing-based proper motions are in reasonable agreement
with each other. An independent interferometric position was obtained
by Fomalont et al.  (1997)\nocite{fgml97}: $\mu_\alpha = -57(7)$ and
$\mu_\delta = -21(11)$~mas~yr$^{-1}$. In right ascension this
measurement agrees well with the timing results but it disagrees with
the Brisken et al. (2003) value. In declination, it is discrepant with
all previous measurements. This pulsar is near the southern horizon
for northern interferometer systems and corrections for ionospheric
refraction are large and rather uncertain. We conclude that the timing
results are probably more reliable, but this remains to be confirmed
by future measurements.

The proper motion estimates in right ascension and declination for
PSR~B2011$+$38 are also discrepant with previous work.  With the large
estimated distance to this pulsar of over 8\,kpc our proper motion
indicates an anomalously high transverse velocity
$>4000$\,km\,s$^{-1}$.  If we compare our position determination with
that obtained by Hobbs et al. (2004) we obtain $\mu_\alpha = -28(17)$
and $\mu_\delta = -30(15)$\,mas\,yr$^{-1}$ using the shorter 7-year
baseline, but more precise position estimate.  These values are
consistent with previous estimates of the pulsar's proper motion and
we therefore conclude that the uncertainty in the catalogue position
was significantly underestimated.

\subsection{Pulsar velocities}

\begin{figure}
\includegraphics[width=9cm]{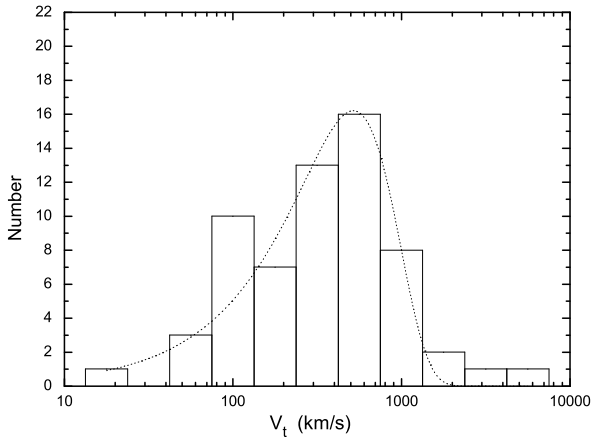}
\caption{Distribution of pulsar transverse velocities for the
pulsars in our sample (values of the pulsar transverse velocity
that have an uncertainty of less than 2000~km~$s^{-1}$ are
included).}\label{fg:histogram}
\end{figure}

The two-dimensional or transverse velocity is given by $V_T =
4.74\mu_{\rm tot} D$~km~s$^{-1}$, where $\mu_{\rm tot}$ is the total
proper motion in mas~yr$^{-1}$ and $D$ the distance in kpc.
Figure~\ref{fg:histogram} shows a histogram of the derived velocities
(listed in Table 2), omitting the anomalously high velocity for
PSR~B2011+38 mentioned above and values which have an uncertainty of
more than 2000~km~s$^{-1}$. This histogram has a similar form to
earlier work. The mean and rms velocities of the sample are
443~km~s$^{-1}$ and 224~km~s$^{-1}$, similar to those given
by Lyne \& Lorimer (1994)\nocite{ll94} (although these authors used
the earlier Taylor \& Cordes\nocite{tc93} electron density model) and
the new results by Hobbs et al. (2005)\nocite{hllk05}.  The velocity
distribution of our sample favours a single-component Gaussian
velocity model, as displayed by the dotted line in
Figure~\ref{fg:histogram}, which is compatible with that of Lyne \&
Lorimer (1994) and Hobbs et al. (2005).

PSR J1721$-$3532 has a very high indicated velocity
(7200~km~s$^{-1}$). While this is only a $3\sigma$ result, such a high
velocity would be unique were it real. However, this result is more
likely to result from a significant under-estimate of position
uncertainties or an over-estimate of the pulsar distance.

\begin{figure}
\includegraphics[width=6.4cm,angle=-90]{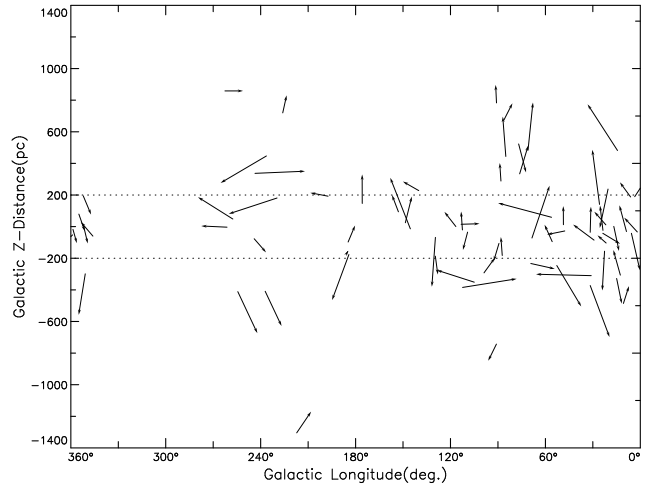}
\caption{Projected directions of the proper motions for the 74
pulsars in our sample relative to the Galactic Plane. The tail end
of the vectors corresponds to the positions given in Table~1 and
the length of the vector represents the total proper motion. The
effects of Galactic rotation and the unknown pulsar radial
velocities have not been taken into account.}\label{fg:direction}
\end{figure}

The motion of these pulsars in the Galactic plane is shown in
Figure~\ref{fg:direction} where the length of the tracks indicate
the pulsar velocities. This figure confirms that most pulsars are
migrating from the Galactic plane \cite{go70} and only a few of
them are moving towards the plane \cite{hla93,las82}. However, for
the majority of these pulsars, more precise proper motions already
exist in the literature. We also note that to make a full study of
the pulsar space velocities, it is essential to consider selection
effects \cite{hp97,acc02}. For instance, the Nanshan telescope
only observed relatively bright pulsars in the the last few years
and so high-velocity pulsars at large $z$-distances may have been
missed \cite{car93,fgw94}. We therefore do not discuss the
velocity distribution in detail and conclude this Section by
discussing the new and improved proper motions from our sample.

\subsection{New and improved proper motions}

\begin{table*}
\caption{New and improved pulsar proper motions.}
  \label{tb:new}
\begin{tabular}{lllllllll}\hline
  PSR J & PSR B & $\mu_\alpha$     & $\mu_\delta$     & Distance   & $V_{1D}^\alpha$ & $V_{1D}^\delta$ & $\tau_c$ & $B_s$ \\
        &       & (mas\,yr$^{-1}$) & (mas\,yr$^{-1}$) & (kpc) & (km\,s$^{-1}$)  & (km\,s$^{-1}$)  & (Myr)    & ($10^{12}$\,G) \\ \hline
  J1707$-$4053 & B1703$-$40 & 3(21)       & 35(61)      & 4.5  & 64(450)     & 750(1300)  & 4.8  & 1.1 \\
  J1740$-$3015 & B1737$-$30 & $-$170(190) & 55(380)     & 2.7  & 2200(2400)  & 710(4900)  & 0.21 & 0.17  \\
  J1741$-$3927 & B1737$-$39 & 20(15)      & $-$6(59)    & 3.2  & 300(230)    & 91(900)    & 4.2  & 1.0 \\
  J1803$-$2137 & B1800$-$21 & 18(30)      & 400(470)    & 3.9  & 330(560)    & 7300(8700) & 0.016 & 4.3 \\
  J1824$-$1945 & B1821$-$19 & $-$12(14)   & $-$100(220) & 4.7  & 270(310)    & 2300(4900) & 0.6  & 1.0\\
  J1829$-$1751 & B1826$-$17 & 22(13)      & $-$150(130) & 4.7  & 490(290)    & 3400(2900) & 0.9  & 1.3\\
  J1835$-$1106 & -----      & 27(46)      & 56(190)     & 2.8  & 360(610)    & 740(2500)  & 0.1  & 1.9\\
  J1836$-$1008 & B1834$-$10 & 18(65)       & 12(220)      & 4.5  & 384(1400)   & 256(4700) & 0.8  & 2.6\\
  J1917$+$1353 & B1915$+$13 & 0(12)       & $-$6(15)    & 4.0  & 0(230)      & 110(280)   &0.4   & 1.2\\ \\

  J0139$+$5814 & B0136$+$57  & $-$21(4) & $-5$(5)    & 2.9 & 290(55) & 69(69) & 0.4 & 1.7 \\
  J0612$+$3721 & B0609$+$37  & 14(5)    & 10(17)     & 0.9 & 60(21) &  43(73) & 79  & 0.1 \\
  J0846$-$3533 & B0844$-$35  & 93(72)   & $-$15(65)  & 0.6 & 260(210) & 43(190)& 11  & 1.4\\
  J1722$-$3207 & B1718$-$32  & $-$1(5)  & $-$40(27)  & 2.4 & 11(57) & 460(300) & 12  & 0.6\\
  J1745$-$3040 & B1742$-$30  & 6(3)     & 4(26)      & 1.9 & 54(27) & 36(230) & 0.5 & 2.0\\
  J1921$+$2153 & B1919$+$21  & 17(4)    & 32(6)      & 1.1 & 89(21) & 167(31) & 16  & 1.4\\
  J2257$+$5909 & B2255$+$58  & 24(6)    & $-$7(5)    & 4.5 & 510(130) & 150(110) & 1.0 & 1.5\\
\hline
\end{tabular}
\end{table*}

In Table~\ref{tb:new} we list the nine pulsars for which we
provide new proper motions (upper part of the table) and a further
seven where our results are more precise than previous
measurements. Besides the proper motions and distances repeated
from Table 2, the 1-D velocities determined from $V_{1D} = 4.74\mu
D$ where $\mu$ is the proper motion in right ascension or
declination, their characteristic age $\tau_c = P/(2\dot{P})$ and
surface dipole magnetic field strength $B_s = 3.2\times 10^{19} (P
\dot{P})^{1/2}$~G are listed. The velocity uncertainties are
determined solely from the uncertainties in the proper motion
estimates and do not include any contribution from the distance
estimates. For PSRs~B0736$-$40 and B0844$-$35 these distances have
decreased by factors of 4.2 and 2.6 respectively with the new
NE2001 distance model compared to the Taylor \& Cordes
(1993)\nocite{tc93} model, significantly reducing the velocity
estimates for these pulsars. Some of the derived velocities are
very large, but in all cases, they are also very uncertain. Proper
motion measurements for these relatively distant pulsars are
difficult and more realistic velocity estimates await improved
measurements.

\section{Conclusion}
We have presented new position and proper motion determinations for 74
pulsars based on five years of timing observations at the Nanshan
radio telescope of Urumqi Observatory. New or improved measurements
are presented for 16 of these pulsars.  It has long been known that
the formal uncertainties given by the standard pulsar timing package
{\sc tempo} are often underestimated.  Using realistic simulations of
pulsar timing noise we show that the uncertainties in position
estimates given by {\sc tempo} should typically be increased by a
factor of 2 -- 3.

The Nanshan radio telescope is continuing to observe these and other
pulsars.  Over the next few years the uncertainties in the positions
and hence the errors in their proper motions should be significantly
reduced.  The data set will also prove valuable in the study of
glitches and timing noise.

\section*{ACKNOWLEDGEMENTS}

 We thank the engineers responsible for maintaining the receiver and
 telescope at Urumqi Observatory of NAO-CAS, and the staff who helped
 with the observations. We also thank the support from the National
 Nature Science Foundation (NNSF) of China under the project 10173020.
 We thank A. Lyne and M. Kramer at Jodrell Bank observatory for
 supplying the TOAs for PSRs~B1133+16 and B0320$+$39 that were used in
 this paper.


\end{document}

%% file: table1.tex
\begin{table*}
\begin{minipage}{160mm}
\caption{Positions of 74 pulsars from Nanshan timing observations and
    previously published positions.  }  \tiny
    \setlength{\tabcolsep}{4.5pt}
\begin{tabular}{lllllllllllllll}
\hline
PSR J   &   PSR B   &   Epoch   &   R.A. (J2000)    &   Dec. (J2000)    &   N   &   T$_{\rm span}$  &   Epoch$^c$   &   R.A.$^c$ (J2000)    &   Dec.$^c$ (J2000)    &   Ref.    \\
    &       &   (MJD)   &   (h~~m~~s)   &   (d~~m~~s)   &       &   (yr)    &   (MJD)   &   (h~~m~~s)   &   (d~~m~~s)   &       \\
\hline
0034$-$0721 &   0031$-$07   &   52288   &   00:34:08.86(5)  &   $-$07:21:52.8(16)   &   107 &   31.8   &   40690   &   00:34:08.88(3)  &   $-$07:21:53.4(7)    &   1   \\
0139+5814   &   0136+57 &   52182   &   01:39:19.742(4) &   +58:14:31.80(4) &   200 &   10.4    &   46573   &   01:39:19.770(3) &   +58:14:31.85(3) &   2   \\
0141+6009   &   0138+59 &   52376   &   01:41:39.92(4)  &   +60:09:32.1(4)  &   156 &   29.0   &   41794   &   01:41:39.947(7) &   +60:09:32.28(5) &   1   \\
0332+5434$^*$   &   0329+54 &   52337   &   03:32:59.391(7) &   +54:34:43.40(8) &   417 &   32.1   &   40105   &   03:32:59.35(1)  &   +54:34:43.3(1)  &   1   \\
0358+5413   &   0355+54 &   52360   &   03:58:53.707(8) &   +54:13:13.8(1)  &   290 &   16.1   &   46573   &   03:58:53.705(4) &   +54:13:13.58(3) &   1   \\  \\
0452$-$1759 &   0450$-$18   &   52394   &   04:52:34.099(4) &   $-$17:59:23.41(7)   &   196 &   15.3   &   46573   &   04:52:34.098(3) &   $-$17:59:23.54(7)   &   1   \\
0454+5543$^*$   &   0450+55 &   52393   &   04:54:07.745(7) &   +55:43:41.45(10)    &   190 &   10.9   &   48416   &   04:54:07.621(3) &   +55:43:41.2(1)  &   1   \\
0528+2200$^*$   &   0525+21 &   52253   &   05:28:52.25(4)  &   +21:59:54(8)    &   147 &   15.7   &   46517   &   05:28:52.308(11)    &   +22:00:01(3)    &   5   \\
0534+2200$^*$   &   0531+21 &   51816   &   05:34:31.8(1)   &   +22:01:10(18)   &   150 &   30.5   &   40675   &   05:34:31.973(5) &   +22:00:52.06(6) &   1   \\
0543+2329   &   0540+23 &   52369   &   05:43:9.66(4)   &   +23:29:08(15)   &   202 &   13.2   &   47555   &   05:43:09.650(3) &   +23:29:06.14(4) &   1   \\  \\
0612+3721   &   0609+37 &   52423   &   06:12:48.675(7) &   +37:21:37.1(3)  &   108 &   17.3   &   46111   &   06:12:48.654(3) &   +37:21:37.0(1)  &   1   \\
0630$-$2834 &   0628$-$28   &   52399   &   06:30:49.418(10)    &   $-$28:34:42.87(18)  &   151 &   22.7   &   40585   &   06:30:49.531(6) &   $-$28:34:43.6(1)    &   1   \\
0738$-$4042$^*$ &   0736$-$40   &   52276   &   07:38:32.305(3) &   $-$40:42:40.10(4)   &   121 &   26.4   &   41331   &   07:38:32.432(3) &   $-$40:42:41.15(4)   &   1   \\
0742$-$2822 &   0740$-$28   &   52388   &   07:42:49.056(10)    &   $-$28:22:43.66(13)  &   171 &   11.0   &   46573   &   07:42:49.073(3) &   $-$28:22:44.0(1)    &   1   \\
0814+7429   &   0809+74 &   52418   &   08:14:59.56(3)  &   +74:29:05.21(16)    &   142 &   11.1   &   48382   &   08:14:59.44(4)  &   +74:29:05.8(1)  &   2   \\  \\
0820$-$1350 &   0818$-$13   &   52398   &   08:20:26.377(4) &   $-$13:50:55.89(14)  &   162 &   31.2   &   46573   &   08:20:26.358(9) &   $-$13:50:55.20(6)   &   1   \\
0826+2637   &   0823+26 &   52398   &   08:26:51.476(9) &   +26:37:22.8(4)  &   406 &   11.0   &   48383   &   08:26:51.310(2) &   +26:37:25.57(7) &   1   \\
0837$-$4135 &   0835$-$41   &   52340   &   08:37:21.1818(17)   &   $-$41:35:14.37(2)   &   93  &   15.2   &   46986   &   08:37:21.173(6) &   $-$41:35:14.29(7)   &   1   \\
0846$-$3533 &   0844$-$35   &   52413   &   08:46:06.05(3)  &   $-$35:33:40.3(5)    &   113 &   24.3   &   43557   &   08:46:05.86(14) &   $-$35:33:39.9(15)   &   1   \\
0922+0638   &   0919+06 &   52264   &   09:22:13.998(14)    &   +06:38:22.1(6)  &   161 &   10.6   &   46573   &   09:22:13.977(3) &   +06:38:21.69(4) &   1   \\  \\
0953+0755   &   0950+08 &   52398   &   09:53:09.278(16)    &   +07:55:35.2(7)  &   183 &   29.8   &   46058   &   09:53:09.304(3) &   +07:55:35.91(4) &   1   \\
1136+1551   &   1133+16 &   52398   &   11:36:03.18(2)  &   +15:51:11.1(3)  &   282 &   29.4   &   42364   &   11:36:03.296(4) &   +15:51:00.7(1)  &   1   \\
1239+2453   &   1237+25 &   52398   &   12:39:40.347(9) &   +24:53:50.38(18)    &   177 &   11.0   &   46058   &   12:39:40.475(3) &   +24:53:49.25(3) &   1   \\
1509+5531   &   1508+55 &   52398   &   15:09:25.635(7) &   +55:31:32.54(6) &   138 &   11.0   &   48383   &   15:09:25.724(9) &   +55:31:33.01(8) &   2   \\
1645$-$0317$^*$ &   1642$-$03   &   52398   &   16:45:02.035(5) &   $-$03:17:58.0(2)    &   247 &   32.2   &   40414   &   16:45:02.045(3) &   $-$03:17:58.4(1)    &   1   \\  \\
1703$-$3241 &   1700$-$32   &   52348   &   17:03:22.530(7) &   $-$32:41:48.0(5)    &   159 &   11.9    &   48000   &   17:03:22.37(12) &   $-$32:41:45(4)  &   1   \\
1705$-$1906 &   1702$-$19   &   52398   &   17:05:36.048(9) &   $-$19:06:38.7(1)    &   147 &   11.1   &   48331   &   17:05:36.108(6) &   $-$19:06:38.5(7)    &   2   \\
1707$-$4053 &   1703$-$40   &   52343   &   17:07:21.747(19)    &   $-$40:53:54.9(6)    &   77  &   10.9    &   48361   &   17:07:21.744(6) &   $-$40:53:55.3(3)    &   3   \\
1709$-$1640$^*$ &   1706$-$16   &   52183   &   17:09:26.436(9) &   $-$16:40:58.0(10)   &   137 &   10.5   &   48367   &   17:09:26.455(2) &   $-$16:40:58.4(3)    &   1   \\
1721$-$3532 &   1718$-$35   &   52396   &   17:21:32.761(8) &   $-$35:32:49.5(5)    &   149 &   11.0  &   48379   &   17:21:32.80(2)  &   $-$35:32:46.6(9)    &   3   \\  \\
1722$-$3207 &   1718$-$32   &   52362   &   17:22:02.953(3) &   $-$32:07:45.5(3)    &   153 &   10.9    &   46956   &   17:22:02.955(4) &   $-$32:07:44.9(3)    &   2   \\
1740$-$3015$^*$ &   1737$-$30   &   52257   &   17:40:33.62(9)  &   $-$30:15:41(4)  &   170 &   8.3    &   49239   &   17:40:33.7(1)   &   $-$30:15:41.9(15)   &   1   \\
1741$-$3927 &   1737$-$39   &   52384   &   17:41:18.081(5) &   $-$39:27:38.0(2)    &   134 &   24.2   &   43558   &   17:41:18.04(3)  &   $-$39:27:37.9(14)   &   1   \\
1745$-$3040 &   1742$-$30   &   52384   &   17:45:56.305(2) &   $-$30:40:23.5(2)    &   153 &   11.0   &   46956   &   17:45:56.299(2) &   $-$30:40:23.6(3)    &   1   \\
1752$-$2806 &   1749$-$28   &   52366   &   17:52:58.693(4) &   $-$28:06:39.0(7)    &   166 &   16.1   &   46483   &   17:52:58.6896(17)   &   $-$28:06:37.3(3)    &   5   \\  \\
1757$-$2421$^*$ &   1754$-$24   &   52390   &   17:57:29.329(6) &   $-$24:22:11(5)  &   142 &   6.8    &   49909   &   17:57:29.3362(14)   &   $-$24:22:07.4(15)   &   5   \\
1803$-$2137$^*$ &   1800$-$21   &   52292   &   18:03:51.37(1)  &   $-$21:37:01(7)  &   172 &   15.7   &   46573   &   18:03:51.35(3)  &   $-$21:37:07.2(5)    &   1   \\
1807$-$0847 &   1804$-$08   &   52261   &   18:07:38.027(4) &   $-$08:47:43.1(2)    &   148 &   15.6   &   46573   &   18:07:38.019(9) &   $-$08:47:43.10(2)   &   1   \\
1818$-$1422 &   1815$-$14   &   52286   &   18:18:23.757(6) &   $-$14:22:37.1(6)    &   144 &   14.5   &   47000   &   18:18:23.79(5)  &   $-$14:22:36(3)  &   1   \\
1820$-$0427 &   1818$-$04   &   52196   &   18:20:52.591(4) &   $-$04:27:37.7(2)    &   153 &   31.7   &   40614   &   18:20:52.621(3) &   $-$04:27:38.5(1)    &   1   \\  \\
1824$-$1945$^*$ &   1821$-$19   &   52267   &   18:24:00.453(6) &   $-$19:45:51.5(14)   &   157 &   15.0   &   49877   &   18:24:00.4555(13)   &   $-$19:45:51.7(3)    &   5   \\
1825$-$0935$^*$ &   1822$-$09   &   52058   &   18:25:30.60(2)  &   $-$09:35:22(3)  &   174 &   10.1   &   48381   &   18:25:30.596(6)     &   $-$09:35:22.8(4)    &   2   \\
1829$-$1751 &   1826$-$17   &   52305   &   18:29:43.143(6) &   $-$17:51:05.2(7)    &   151 &   14.7   &   46944   &   18:29:43.121(12)    &   $-$17:51:02.9(18)   &   1   \\
1832$-$0827 &   1829$-$08   &   52261   &   18:32:37.016(5) &   $-$08:27:03.4(3)    &   172 &   14.4    &   47000   &   18:32:37.024(7) &   $-$08:27:03.7(3)    &   1   \\
1833$-$0827 &   1830$-$08   &   52306   &   18:33:40.286(4) &   $-$08:27:31.5(2)    &   134 &   11.7   &   48041   &   18:33:40.32(2)  &   $-$08:27:30.7(6)    &   1   \\  \\
1835$-$1106$^*$ &   ---  &   51884   &   18:35:18.30(2)  &   $-$11:06:14.7(12)   &   149 &   6.7    &   49446   &   18:35:18.287(2) &   $-$11:06:15.1(2)    &   4   \\
1836$-$1008 &   1834$-$10   &   52261   &   18:36:53.919(6) &   $-$10:08:09.0(4)    &   147 &   22.9   &   43893   &   18:36:53.9(1)   &   $-$10:08:09(5)  &   1   \\
1840+5640   &   1839+56 &   52285   &   18:40:44.57(3)  &   +56:40:55.3(2)  &   136 &   10.7   &   48381   &   18:40:44.59(5)  &   +56:40:55.6(4)  &   2   \\
1847$-$0402 &   1844$-$04   &   52286   &   18:47:22.835(6) &   $-$04:02:14.1(3)    &   137 &   15.0  &   46956   &   18:47:22.834(14)    &   $-$04:02:14.2(5)    &   1   \\
1848$-$0123$^*$ &   1845$-$01   &   52261   &   18:48:23.588(8) &   $-$01:23:58.1(3)    &   144 &   15.0   &   46987   &   18:48:23.60(2)  &   $-$01:23:58.2(7)    &   1   \\  \\
1900$-$2600 &   1857$-$26   &   52298   &   19:00:47.56(2)  &   $-$26:00:44(1)  &   140 &   10.7   &   46573   &   19:00:47.596(10)    &   $-$26:00:43.1(3)    &   1   \\
1913$-$0440 &   1911$-$04   &   52385   &   19:13:54.182(4) &   $-$04:40:47.98(17)  &   131 &   32.2    &   41902   &   19:13:54.18(1)  &   $-$04:40:47.6(4)    &   1   \\
1917+1353   &   1915+13 &   52298   &   19:17:39.784(8) &   +13:53:56.99(13)    &   121 &   10.7   &   48382   &   19:17:39.784(2) &   +13:53:57.06(9) &   2   \\
1921+2153   &   1919+21 &   52311   &   19:21:44.832(8) &   +21:53:02.7(1)  &   87  &   31.8   &   42084   &   19:21:44.798(3) &   +21:53:01.83(8) &   1   \\
1932+1059   &   1929+10 &   52385   &   19:32:13.957(6) &   +10:59:32.8(1)  &   391 &   11.0   &   48381   &   19:32:13.900(2) &   +10:59:31.99(7) &   2   \\  \\
1935+1616   &   1933+16 &   52286   &   19:35:47.820(5) &   +16:16:40.1(1)  &   133 &   27.4   &   40213   &   19:35:47.835(1) &   +16:16:40.59(2) &   1   \\
1946+1805   &   1944+17 &   52287   &   19:46:53.04(1)  &   +18:05:41.6(3)  &   139 &   27.3   &   42320   &   19:46:53.043(4) &   +18:05:41.59(9) &   1   \\
1948+3540$^*$   &   1946+35 &   52285   &   19:48:24.999(8) &   +35:40:11.07(11)    &   131 &   27.6   &   42221   &   19:48:25.037(2) &   +35:40:11.28(2) &   1   \\
1955+5059   &   1953+50 &   52268   &   19:55:18.713(7) &   +50:59:56.03(7) &   153 &   22.0   &   44240   &   19:55:18.90(6)  &   +50:59:54.2(6)  &   1   \\
2002+4050   &   2000+40 &   52291   &   20:02:44.0267(9)    &   +40:50:53.96(10)    &   153 &   16.9   &   46105   &   20:02:44.04(3)  &   +40:50:54.7(3)  &   1   \\  \\
2013+3845   &   2011+38 &   52285   &   20:13:10.35(1)  &   +38:45:43.1(1)  &   144 &   17.0  &   46075   &   20:13:10.49(3)  &   +38:45:44.8(3)  &   1   \\
2018+2839   &   2016+28 &   52285   &   20:18:03.817(5) &   +28:39:54.28(9) &   138 &   33.4   &   40105   &   20:18:03.85(3)  &   +28:39:54.264(2)    &   1   \\
2022+2854   &   2020+28 &   52373   &   20:22:37.061(5) &   +28:54:23.05(12)    &   380 &   27.4   &   42370   &   20:22:37.079(3) &   +28:54:23.45(3) &   1   \\
2022+5154   &   2021+51 &   52398   &   20:22:49.859(6) &   +51:54:50.49(5) &   455 &   11.0  &   40614   &   20:22:49.900(2) &   +51:54:50.060(2)    &   1   \\
2048$-$1616 &   2045$-$16   &   52281   &   20:48:35.53(4)  &   $-$16:16:41(2)  &   120 &   15.6   &   46573   &   20:48:35.472(4) &   $-$16:16:44.45(8)   &   1   \\  \\
2108+4441   &   2106+44 &   52183   &   21:08:20.48(1)  &   +44:41:48.9(1)  &   150 &   10.4    &   48383   &   21:08:20.48(1)  &   +44:41:48.8(1)  &   2   \\
2113+4644   &   2111+46 &   52284   &   21:13:24.34(2)  &   +46:44:08.8(2)  &   144 &   10.7   &   48382   &   21:13:24.295(14)    &   +46:44:08.68(11)    &   2   \\
2157+4017   &   2154+40 &   52284   &   21:57:1.85(2)   &   +40:17:46.0(1)  &   160 &   10.7   &   48382   &   21:57:01.821(13)    &   +40:17:45.9(1)  &   2   \\
2219+4754   &   2217+47 &   52264   &   22:19:48.106(7) &   +47:54:53.68(7) &   156 &   10.6   &   48382   &   22:19:48.136(4) &   +47:54:53.83(4) &   2   \\
2257+5909   &   2255+58 &   52284   &   22:57:57.745(7) &   +59:09:14.87(5) &   162 &   10.6   &   48419   &   22:57:57.711(4) &   +59:09:14.95(3) &   2   \\  \\
2313+4253   &   2310+42 &   52373   &   23:13:08.618(7) &   +42:53:13.16(9) &   158 &   15.9   &   46573   &   23:13:08.571(6) &   +42:53:12.98(3) &   1   \\
2321+6024   &   2319+60 &   52261   &   23:21:55.18(1)  &   +60:24:30.7(1)  &   154 &   10.6   &   48383   &   23:21:55.19(4)  &   +60:24:30.7(3)  &   2   \\
2326+6113   &   2324+60 &   52291   &   23:26:58.672(7) &   +61:13:36.42(5) &   151 &   10.6   &   48416   &   23:26:58.704(5) &   +61:13:36.50(3) &   1   \\
2354+6155   &   2351+61 &   52285   &   23:54:04.76(1)  &   +61:55:46.79(8) &   147 &   10.7   &   48382   &   23:54:04.710(17)    &   +61:55:46.8(1)  &   2   \\
\hline
\end{tabular}
$^*$ Quoted errors are three times the formal {\sc tempo} values (see text); others are two times the formal {\sc tempo}
values.

References for the previously published data are: (1) Taylor,
Manchester \& Lyne, 1993; (2) Arzoumanian et al., 1994; (3)
Johnston et al., 1995; (4) Manchester et al., 1996. (5) Hobbs et
al., 2004.
\end{minipage}
\end{table*}
\nocite{tml93,antt94,jml+95,mld+96}

%% file: table2.tex
\begin{table*}
\begin{minipage}{175mm}
\caption{Comparison of pulsar proper motions derived from Nanshan
  timing results with previously published values. Proper motions are given as mas~yr$^{-1}$ on the sky, i.e., $\mu_\alpha=\dot{\alpha}\;\cos\delta$.}
\tiny \setlength{\tabcolsep}{4.2pt}
\begin{tabular}{lllllllllllllll}
\hline
PSR J   &   PSR B   &   Dist    &   $\mu_{\alpha}^c$    &   $\mu_\delta^c$  &   $\mu_{tot}^c$   &   $\mu_{\alpha}^T$    &   ${\mu_\delta}^T$    &   $\mu_{\alpha}^I$    &   ${\mu_\delta}^I$    &   Ref.    &   $\epsilon_\alpha$   &   $\epsilon_\delta$   &   $V_t$   \\
    &       &   (kpc)   &   (mas~yr$^{-1}$) &   (mas~yr$^{-1}$) &   (mas~yr$^{-1}$) &   (mas~yr$^{-1}$) &   (mas~yr$^{-1}$) &   (mas~yr$^{-1}$) &   (mas~yr$^{-1}$) &       &       &       &   (km~s$^{-1}$)   \\
\hline
0034$-$0721 &   0031$-$07   &   0.4     &   $-$7(27)    &   18(56)  &   19(53)  &   $-$16(26)   &   17(53)  &   $-$102(74)  &   $-$105(78)  &   6   &   0.2     &   0.0     &   38(100) \\
0139+5814   &   0136+57 &   2.9     &   $-$21(4)    &   $-$5(5) &   21(4)   &   --- &   --- &   $-$11(5)    &   $-$19(5)    &   1   &   1.1     &   1.4     &   290(52) \\
0141+6009   &   0138+59 &   2.2     &   $-$29(34)   &   $-$12(23)   &   32(33)  &   10(14)  &   $-$5(15)    &   --- &   --- &   --- &   0.8     &   0.2     &   330(340)    \\
0332+5434   &   0329+54 &   1.1     &   12(3)   &   5(4)    &   12(3)   &   12(4)   &   $-$14(7)    &   17.0(3) &   $-$9.5(4)   &   8   &   1.5     &   3.3     &   63(15)  \\
0358+5413   &   0355+54 &   1.1     &   1(5)    &   19(7)   &   19(7)   &   8(3)    &   19(6)   &   9.2(2)  &   8.2(4)  &   9   &   1.6     &   1.4     &   99(37)  \\  \\
0452$-$1759 &   0450$-$18   &   2.4     &   1(4)    &   8(7)    &   8(6)    &   10(4)   &   3(6)    &   12(8)   &   18(15)  &   3   &   1.1     &   0.4     &   94(70)  \\
0454+5543   &   0450+55 &   0.7     &   64(4)   &   14(9)   &   66(4)   &   48(6)   &   $-$13(12)   &   52(6)   &   $-$17(2)    &   1   &   1.3     &   2.9     &    210(10)    \\
0528+2200   &   0525+21 &   1.6     &   $-$24(20)   &   $-$520(570) &   520(570)    &   $-$38(15)   &   200(300)    &   $-$20(19)   &   $-$7(9) &   1   &   0.1     &   0.9     &   5200(2800)  \\
0534+2200   &   0531+21 &   1.7     &   $-$75(44)   &   570(590)    &   580(580)    &   --- &   --- &   $-$16(11)   &   $-$2(8) &   4   &   1.1     &   1.0     &   4800(4800)  \\
0543+2329   &   0540+23 &   2.1     &   6(42)   &   140(1100)   &   140(1100)   &   22(32)  &   $-$400(1100)    &   19(7)   &   12(8)   &   1   &   0.3     &   0.1     &   1400(11000) \\  \\
0612+3721   &   0609+37 &   0.9     &   14(5)   &   10(17)  &   17(11)  &   5(5)    &   6(22)   &   --- &   --- &   --- &   0.9     &   0.1     &   71(44)  \\
0630$-$2834 &   0628$-$28   &   1.5     &   $-$46(5)    &   23(7)   &   51(5)   &   $-$51(7)    &   17(8)   &   $-$44.6(9)  &   20(2)   &   2   &   0.3     &   0.3     &   350(37) \\
0738$-$4042 &   0736$-$40   &   2.6     &   $-$48(2)    &   35(2)   &   59(2)   &   $-$60(10)   &   40(10)  &   $-$14.0(12) &   13(2)   &   7   &   11.9    &   5.5     &   740(18) \\
0742$-$2822 &   0740$-$28   &   2.1     &   $-$14(9)    &   23(12)  &   27(11)  &   18(20)  &   $-$49(21)   &   $-$29(2)    &   4(2)    &   3   &   1.3     &   1.3     &   260(110)    \\
0814+7429   &   0809+74 &   0.4     &   31(13)  &   $-$52(19)   &   61(18)  &   --- &   --- &   24.02(9)    &   $-$44.0(4)  &   8   &   0.5     &   0.4     &   120(36) \\  \\
0820$-$1350 &   0818$-$13   &   2.0     &   18(9)   &   $-$44(9)    &   47(9)   &   21(3)   &   $-$41(5)    &   9(9)    &   $-$47(6)    &   3   &   0.3     &   0.2     &   440(85) \\
0826+2637   &   0823+26 &   0.4     &   67(4)   &   $-$82(14)   &   110(11) &   65(3)   &   $-$108(8)   &   61(3)   &   $-$90(2)    &   4   &   0.9     &   0.5     &   180(18) \\
0837$-$4135 &   0835$-$41   &   1.1     &   7(5)    &   $-$6(5) &   9(5)    &   --- &   --- &   $-$2.3(18)  &   $-$18(3)    &   2   &   1.4     &   1.5     &   44(24)  \\
0846$-$3533 &   0844$-$35   &   0.6     &   93(72)  &   $-$15(65)   &   94(72)  &   54(70)  &   $-$8(85)    &   --- &   --- &   --- &   0.3     &   0.0     &   250(190)    \\
0922+0638   &   0919+06 &   1.2     &   21(14)  &   28(34)  &   34(31)  &   --- &   --- &   18.8(9) &   86.4(7) &   7   &   0.1     &   1.6     &   200(170)    \\  \\
0953+0755   &   0950+08 &   0.3     &   $-$22(14)   &   $-$39(41)   &   45(36)  &   $-$3(3) &   26(7)   &   $-$2.09(8)  &   29.46(7)    &   8   &   1.4     &   1.7     &   55(45)  \\
1136+1551   &   1133+16 &   0.4     &   $-$61(11)   &   380(10) &   380(10) &   $-$69(3)    &   327(5)  &   $-$74.0(4)  &   368.1(3)    &   8   &   1.2     &   1.0     &   650(17) \\
1239+2453   &   1237+25 &   0.9     &   $-$100(7)   &   65(10)  &   120(8)  &   $-$108.8(14)    &   50(3)   &   $-$104.5(11)    &   49.4(14)    &   7   &   0.4     &   1.4     &   490(34) \\
1509+5531   &   1508+55 &   1.0     &   $-$75(9)    &   $-$49(9)    &   89(9)   &   $-$86(12)   &   $-$65(11)   &   $-$70.6(16) &   $-$68.8(12) &   7   &   0.4     &   1.9     &   420(42) \\
1645$-$0317 &   1642$-$03   &   1.1     &   $-$4(3) &   10(8)   &   11(7)   &   --- &   --- &   $-$3.7(15)  &   30.5(16)    &   2   &   0.1     &   2.2     &   57(29)  \\  \\
1703$-$3241 &   1700$-$32   &   2.3     &   170(130)    &   $-$240(300) &   290(250)    &   $-$26(17)   &   62(104) &   --- &   --- &   --- &   1.4     &   0.7     &   3200(2800)  \\
1705$-$1906 &   1702$-$19   &   0.9     &   $-$77(14)   &   $-$15(120)  &   78(26)  &   $-$78(9)    &   $-$116(82)  &   --- &   --- &   --- &   0.0     &   0.5     &   330(110)    \\
1707$-$4053 &   1703$-$40   &   4.5     &   3(21)   &   35(61)  &   35(61)  &   --- &   --- &   --- &   --- &   --- &   --- &   --- &   760(1300)   \\
1709$-$1640 &   1706$-$16   &   0.8     &   $-$9(4) &   13(34)  &   16(28)  &   --- &   --- &   3(9)    &   0(14)   &   3   &   0.9     &   0.3     &   64(77)  \\
1721$-$3532 &   1718$-$35   &   5.6     &   $-$43(24)   &   $-$270(93)  &   270(91) &   $-$6(105)   &   $-$600(600) &   --- &   --- &   --- &   0.3     &   0.5     &   7200(2400)  \\  \\
1722$-$3207 &   1718$-$32   &   2.4     &   $-$1(5) &   $-$40(27)   &   40(27)  &   $-$5(8) &   5(54)   &   --- &   --- &   --- &   0.3     &   0.6     &   450(300)    \\
1740$-$3015 &   1737$-$30   &   2.7     &   $-$170(190) &   55(380) &   180(210)    &   --- &   --- &   --- &   --- &   --- &   --- &   --- &   2400(2700)  \\
1741$-$3927 &   1737$-$39   &   3.2     &   20(15)  &   $-$6(59)    &   21(22)  &   --- &   --- &   --- &   --- &   --- &   --- &   --- &   320(340)    \\
1745$-$3040 &   1742$-$30   &   1.9     &   6(3)    &   4(26)   &   7(15)   &   13(4)   &   81(32)  &   --- &   --- &   --- &   1.1     &   1.3     &   65(140) \\
1752$-$2806 &   1749$-$28   &   1.2     &   3(4)    &   $-$110(47)  &   110(48) &   1.1(23) &   44(27)  &   $-$4(6) &   $-$5(5) &   3   &   0.7     &   1.9     &   630(280)    \\  \\
1757$-$2421 &   1754$-$24   &   4.4     &   $-$15(12)   &   $-$520(760) &   520(760)    &   $-$20(6)    &   $-$500(400) &   --- &   --- &   --- &   0.3     &   0.0     &   11000(11000)    \\
1803$-$2137 &   1800$-$21   &   3.9     &   18(30)  &   400(470)    &   400(470)    &   --- &   --- &   --- &   --- &   --- &   --- &   --- &   7400(5700)  \\
1807$-$0847 &   1804$-$08   &   2.7     &   7(9)    &   1(19)   &   7(10)   &   $-$2.3(12)  &   $-$4(5) &   $-$5(4) &   1(4)    &   3   &   0.2     &   0.0     &   97(120) \\
1818$-$1422 &   1815$-$14   &   7.2     &   $-$29(51)   &   $-$87(210)  &   92(202) &   $-$6(11)    &   $-$78(65)   &   --- &   --- &   --- &   0.4     &   0.0     &   3100(6800)  \\
1820$-$0427 &   1818$-$04   &   1.9     &   $-$14(2)    &   26(7)   &   29(6)   &   $-$10(3)    &   9(9)    &   3(3)    &   27(3)   &   4   &   3.1     &   0.1     &   270(57) \\  \\
1824$-$1945 &   1821$-$19   &   4.7     &   $-$12(14)   &   $-$100(220) &   100(220)    &   --- &   --- &   --- &   --- &   --- &   --- &   --- &   2300(3300)  \\
1825$-$0935 &   1822$-$09   &   0.9     &   11(23)  &   39(180) &   41(170) &   --- &   --- &   $-$13(11)   &   $-$9(5) &   3   &   0.7     &   0.3     &   170(710)    \\
1829$-$1751 &   1826$-$17   &   4.7     &   22(13)  &   $-$150(130) &   160(130)    &   --- &   --- &   --- &   --- &   --- &   --- &   --- &   3500(2900)  \\
1832$-$0827 &   1829$-$08   &   4.9     &   $-$8(9) &   16(20)  &   18(18)  &   $-$4(4) &   20(15)  &   --- &   --- &   --- &   0.3     &   0.1     &   410(410)    \\
1833$-$0827 &   1830$-$08   &   4.7     &   $-$38(26)   &   $-$72(55)   &   82(50)  &   $-$37(4)  &   0(15)   &   --- &   --- &   --- &   0.0     &   1.0     &   1800(1100)  \\  \\
1835$-$1106 &   ------$-$   &   2.8     &   27(46)  &   56(190) &   62(170) &   --- &   --- &   --- &   --- &   --- &   --- &   --- &   830(1600)   \\
1836$-$1008 &   1834$-$10   &   4.5     &   18(65)  &   12(220) &   22(140) &   --- &   --- &   --- &   --- &   --- &   --- &   --- &   460(2900)   \\
1840+5640   &   1839+56 &   1.7     &   $-$17(45)   &   $-$29(43)   &   34(44)  &   $-$29(11)   &   $-$33(12)   &   $-$30(4)    &   $-$21(2)  &   1   &   0.3     &   0.2     &   270(350)    \\
1847$-$0402 &   1844$-$04   &   3.3     &   1(15)   &   5(38)   &   5(38)   &   $-$1(5) &   $-$9(19)    &   --- &   --- &   --- &   0.1     &   0.3     &   83(590) \\
1848$-$0123 &   1845$-$01   &   4.0     &   $-$12(22)   &   7(53)   &   14(33)  &   2(6)    &   $-$41(19)   &   --- &   --- &   --- &   0.5     &   0.7     &   270(590)    \\  \\
1900$-$2600 &   1857$-$26   &   2.0     &   $-$30(17)   &   $-$35(67)   &   47(52)  &   $-$31(13)   &   7(105)  &   $-$19.9(3)  &   $-$47(1)    &   10  &   0.6     &   0.2     &   440(490)    \\
1913$-$0440 &   1911$-$04   &   2.8     &   1(6)    &   $-$12(15)   &   13(15)  &   7(3)    &   $-$26(7)    &   7(13)   &   $-$5(9) &   1   &   0.7     &   0.6     &   160(200)    \\
1917+1353   &   1915+13 &   4.0     &   0(12)   &   $-$6(15)    &   6(15)   &   --- &   --- &   --- &   --- &   --- &   --- &   --- &   110(280)    \\
1921+2153   &   1919+21 &   1.1     &   17(4)   &   32(6)   &   37(6)   &   23(6)   &   29(12)  &   9(24)   &   40(10)  &   3   &   0.6     &   0.2     &   190(29) \\
1932+1059   &   1929+10 &   0.4     &   76(8)   &   71(12)  &   100(10) &   93.0(18)    &   45(4)   &   94.09(11)   &   42.99(16)   &   9   &   2.2     &   2.3     &   180(17) \\  \\
1935+1616   &   1933+16 &   5.6     &   $-$6(2) &   $-$15(3)    &   16(3)   &   1.1(4)  &   $-$16.3(7)  &   $-$1(3) &   $-$13(3)    &   3   &   2.7     &   0.3     &   440(87) \\
1946+1805   &   1944+17 &   1.4     &   0(7)    &   2(10)   &   2(10)   &   7(15)   &   $-$9(18)    &   1(5)    &   $-$9(4) &   4   &   0.1     &   0.8     &   10(64)  \\
1948+3540   &   1946+35 &   5.8     &   $-$17(4)    &   $-$7(4) &   19(4)   &   $-$14(4)    &   $-$3(4) &   $-$12.6(6)  &   0.7(6)  &   7   &   1.0     &   1.6     &   510(76) \\
1955+5059   &   1953+50 &   2.2     &   $-$79(26)   &   82(27)  &   110(27) &   $-$30(3)    &   59(30)  &   $-$23(5)    &   54(5)   &   1   &   1.8     &   0.8     &   1200(280)   \\
2002+4050   &   2000+40 &   5.9     &   $-$9(21)    &   $-$44(19)   &   45(19)  &   $-$17(9)    &   $-$11(9)    &   --- &   --- &   --- &   0.3     &   1.2     &   1300(530)   \\  \\
2013+3845   &   2011+38 &   8.4     &   $-$97(22)   &   $-$97(19)   &   140(20) &   $-$37(9)    &   $-$35(11)   &   $-$32.1(17) &   $-$24.9(23) &   2   &   2.8     &   3.4     &   5500(810)   \\
2018+2839   &   2016+28 &   1.0     &   $-$13(12)   &   1(3)    &   13(12)  &   $-$2.5(11)  &   $-$4.6(15)  &   2.6(2)  &   6.2(4)  &   8   &   1.3     &   1.7     &   61(55)  \\
2022+2854   &   2020+28 &   2.7     &   $-$9(3) &   $-$15(5)    &   17(4)   &   --- &   --- &   $-$4.4(5)   &   $-$23.6(3)  &   8   &   1.3     &   1.8     &   220(54) \\
2022+5154   &   2021+51 &   2.0     &   $-$12(2)    &   14(2)   &   18(2)   &   $-$5.7(10)  &   $-$10.7(10) &   $-$5.23(17) &   11.5(3) &   8   &   2.7     &   1.2     &   1700(17)    \\
2048$-$1616 &   2045$-$16   &   0.6     &   54(41)  &   220(150)    &   230(150)    &   141(14) &   $-$105(50)  &   117(5)  &   $-$5(5) &   3   &   1.4     &   1.5     &   600(380)    \\  \\
2108+4441   &   2106+44 &   5.0     &   4(17)   &   10(17)  &   11(17)  &   8(12)   &   $-$10(11)   &   --- &   --- &   --- &   0.1     &   0.7     &   260(400)    \\
2113+4644   &   2111+46 &   4.5     &   42(21)  &   14(20)  &   45(21)  &   9(15)   &   $-$1(16)    &   --- &   --- &   --- &   0.9     &   0.4     &   960(450)    \\
2157+4017   &   2154+40 &   3.8     &   35(19)  &   11(19)  &   36(19)  &   0(17)   &   7(15)   &   17.8(8) &   2.8(10) &   2   &   1.0     &   0.1     &   650(330)    \\
2219+4754   &   2217+47 &   2.2     &   $-$28(8)    &   $-$15(8)    &   32(8)   &   $-$14(3)    &   $-$18(3)    &   $-$12(8)    &   $-$30(6)    &   4   &   1.3     &   0.3     &   335(82) \\
2257+5909   &   2255+58 &   4.5     &   24(6)   &   $-$7(5) &   25(6)   &   28(7)   &   $-$8(6) &   --- &   --- &   --- &   0.3     &   0.0     &   550(120)    \\  \\
2313+4253   &   2310+42 &   1.3     &   33(6)   &   11(6)   &   35(6)   &   21.2(15)    &   6.3(15) &   $-$42(30)   &   21(29)  &   5   &   1.5     &   0.7     &   210(37) \\
2321+6024   &   2319+60 &   3.0     &   $-$7(30)    &   4(30)   &   8(30)   &   $-$17(22)   &   $-$7(19)    &   --- &   --- &   --- &   0.2     &   0.2     &   110(430)    \\
2326+6113   &   2324+60 &   4.9     &   $-$22(6)    &   $-$7(5) &   23(6)   &   $-$17(5)    &   $-$9(5) &   --- &   --- &   --- &   0.4     &   0.2     &   530(130)    \\
2354+6155   &   2351+61 &   3.4     &   30(14)  &   1(14)   &   30(14)  &   18(8)   &   $-$1(6) &   22(3)   &   6(2)    &   1   &   0.5     &   0.3     &   490(220)    \\
\hline
\end{tabular}

References to previously published values are: (1) Harrison, Lyne \&
Anderson, 1993; (2) Brisken, 2001; (3) Fomalont et al., 1997; (4)
Lyne, Anderson \& Salter, 1982; (5) Fomalont et al., 1992; (6) Han
\& Tian, 1999; (7) Brisken et al. (2003); (8) Brisken et al., 2002;
(9) Chatterjee et al., 2004; (10) Fomalont et al., 1999.
\end{minipage}
\end{table*}
\nocite{hla93,bri01,fgml97,las82,fgl+92,ht99,bfg+03,bbgt02,ccv+04,fgbc99}

%% file: ver15.bbl
\begin{thebibliography}{{{D'Alessandro}, {Deshpande} \& {McCulloch} }{1997}}

\bibitem[\protect\citename{Arzoumanian, Chernoff \& Cordes }{2002}]{acc02}
Arzoumanian~Z., Chernoff~D.~F., Cordes~J.~M., 2002, ApJ, 568, 289

\bibitem[\protect\citename{Arzoumanian {\rm et~al. }}{1994}]{antt94}
Arzoumanian~Z., Nice~D.~J., Taylor~J.~H., Thorsett~S.~E., 1994,
ApJ, 422, 671

\bibitem[\protect\citename{Boynton {\rm et~al. }}{1972}]{bgh+72}
Boynton~P.~E., Groth~E.~J., Hutchinson~D.~P., Nanos~G.~P.,
Partridge~R.~B.,
  Wilkinson~D.~T., 1972, ApJ, 175, 217

\bibitem[\protect\citename{Brisken }{2001}]{bri01}
Brisken~W.~F., 2001, {\rm PhD thesis}, Princeton University

\bibitem[\protect\citename{Brisken {\rm et~al. }}{2002}]{bbgt02}
Brisken~W.~F., Benson~J.~M., Goss~W.~M., Thorsett~S.~E., 2002,
ApJ, 571, 906

\bibitem[\protect\citename{{Brisken} {\rm et~al. }}{2003}]{bfg+03}
{Brisken}~W.~F., {Fruchter}~A.~S., {Goss}~W.~M.,
{Herrnstein}~R.~M.,
  {Thorsett}~S.~E., 2003, AJ, 126, 3090

\bibitem[\protect\citename{Caraveo }{1993}]{car93}
Caraveo~P.~A., 1993, ApJ, 415, L111

\bibitem[\protect\citename{{Chatterjee} {\rm et~al. }}{2004}]{ccv+04}
{Chatterjee}~S., {Cordes}~J.~M., {Vlemmings}~W.~H.~T.,
{Arzoumanian}~Z.,
  {Goss}~W.~M., {Lazio}~T.~J.~W., 2004, ApJ, 604, 339

\bibitem[\protect\citename{{Cordes} \& Lazio }{2002}]{cl02}
{Cordes}~J.~M., Lazio~T.~J.~W., 2002,
http://xxx.lanl.gov/abs/astro-ph/0207156

\bibitem[\protect\citename{{D'Alessandro}, {Deshpande} \& {McCulloch}
  }{1997}]{ddm97}
{D'Alessandro}~F., {Deshpande}~A.~A., {McCulloch}~P.~M., 1997,
JApA, 18, 5

\bibitem[\protect\citename{Downs \& Reichley }{1983}]{dr83}
Downs~G.~S., Reichley~P.~E., 1983, ApJS, 53, 169

\bibitem[\protect\citename{Fomalont {\rm et~al. }}{1992}]{fgl+92}
Fomalont~E.~B., Goss~W.~M., Lyne~A.~G., Manchester~R.~N.,
Justtanont~K., 1992,
  MNRAS, 258, 497

\bibitem[\protect\citename{Fomalont {\rm et~al. }}{1997}]{fgml97}
Fomalont~E.~B., Goss~W.~M., Manchester~R.~N., Lyne~A.~G., 1997,
MNRAS, 286, 81

\bibitem[\protect\citename{Fomalont {\rm et~al. }}{1999}]{fgbc99}
Fomalont~E.~B., Goss~W.~M., Beasley~A.~J., Chatterjee~S., 1999,
AJ, 117, 3025

\bibitem[\protect\citename{Frail, Goss \& Whiteoak }{1994}]{fgw94}
Frail~D.~A., Goss~W.~M., Whiteoak~J. B.~Z., 1994, ApJ, 437, 781

\bibitem[\protect\citename{Groth }{1975}]{gro75b}
Groth~E.~J., 1975, ApJS, 29, 443

\bibitem[\protect\citename{Gunn \& Ostriker }{1970}]{go70}
Gunn~J.~E., Ostriker~J.~P., 1970, ApJ, 160, 979

\bibitem[\protect\citename{{Han} \& {Tian} }{1999}]{ht99}
{Han}~J.~L., {Tian}~W.~W., 1999, A\&AS, 136, 571

\bibitem[\protect\citename{Hansen \& Phinney }{1997}]{hp97}
Hansen~B., Phinney~E.~S., 1997, MNRAS, 291, 569

\bibitem[\protect\citename{Harrison, Lyne \& Anderson }{1993}]{hla93}
Harrison~P.~A., Lyne~A.~G., Anderson~B., 1993, MNRAS, 261, 113

\bibitem[\protect\citename{Hobbs, Edwards \& Manchester }{2005}]{hem05}
Hobbs~G.~B., Edwards~R.~T., Manchester~R.~N., 2005, PASA, In
preparation

\bibitem[\protect\citename{{Hobbs} {\rm et~al. }}{2004}]{hlk+04}
{Hobbs}~G., {Lyne}~A.~G., {Kramer}~M., {Martin}~C.~E.,
{Jordan}~C.~A., 2004,
  MNRAS, 353, 1311

\bibitem[\protect\citename{Hobbs {\rm et~al. }}{2005}]{hllk05}
Hobbs~G.~B., Lorimer~D.~R., Lyne~A.~G., Kramer~M., 2005, MNRAS,
Accepted

\bibitem[\protect\citename{{Hotan}, {van Straten} \& {Manchester}
  }{2004}]{hvm04}
{Hotan}~A.~W., {van Straten}~W., {Manchester}~R.~N., 2004, PASA,
21, 302

\bibitem[\protect\citename{Johnston {\rm et~al. }}{1995}]{jml+95}
Johnston~S., Manchester~R.~N., Lyne~A.~G., Kaspi~V.~M.,
D'Amico~N., 1995, A\&A,
  293, 795

\bibitem[\protect\citename{Lyne \& Lorimer }{1994}]{ll94}
Lyne~A.~G., Lorimer~D.~R., 1994, Nature, 369, 127

\bibitem[\protect\citename{Lyne, Anderson \& Salter }{1982}]{las82}
Lyne~A.~G., Anderson~B., Salter~M.~J., 1982, MNRAS, 201, 503

\bibitem[\protect\citename{Manchester {\rm et~al. }}{1996}]{mld+96}
Manchester~R.~N. {\rm et~al.}, 1996, MNRAS, 279, 1235

\bibitem[\protect\citename{Manchester, Taylor \& Van }{1974}]{mtv74}
Manchester~R.~N., Taylor~J.~H., Van~Y.-Y., 1974, ApJ, 189, L119

\bibitem[\protect\citename{Siegman, Manchester \& Durdin }{1993}]{smd93}
Siegman~B.~C., Manchester~R.~N., Durdin~J.~M., 1993, MNRAS, 262,
449

\bibitem[\protect\citename{Taylor \& Cordes }{1993}]{tc93}
Taylor~J.~H., Cordes~J.~M., 1993, ApJ, 411, 674

\bibitem[\protect\citename{Taylor, Manchester \& Lyne }{1993}]{tml93}
Taylor~J.~H., Manchester~R.~N., Lyne~A.~G., 1993, ApJS, 88, 529

\bibitem[\protect\citename{{Toscano} {\rm et~al. }}{1999}]{tsb+99}
{Toscano}~M., {Sandhu}~J.~S., {Bailes}~M., {Manchester}~R.~N.,
{Britton}~M.~C.,
  {Kulkarni}~S.~R., {Anderson}~S.~B., {Stappers}~B.~W., 1999, MNRAS, 307, 925

\bibitem[\protect\citename{Wang {\rm et~al. }}{2001}]{wmz+01}
Wang~N., Manchester~R.~N., Zhang~J., Wu~X.~J., Yusup~A.,
Lyne~A.~G.,
  Cheng~K.~S., Chen~M.~Z., 2001, MNRAS, 328, 855

\bibitem[\protect\citename{{Zou} {\rm et~al. }}{2004}]{zww+04}
{Zou}~W.~Z., {Wang}~N., {Wang}~H.~X., {Manchester}~R.~N.,
{Wu}~X.~J.,
  {Zhang}~J., 2004, MNRAS, 354, 811

\end{thebibliography}
